\documentclass[twocolumn,english,showpacs]{revtex4}\newcommand{\JHEPonly}[1]{}
\usepackage{graphicx}
\usepackage{amsmath,amssymb,bm}

 \def\imo{i}

\def\imo{i}

\def\be{\begin{equation}}
\def\ee{\end{equation}}
\def\bea{\begin{eqnarray}}
\def\eea{\end{eqnarray}}

\def\imo{i}

\def\K{{\cal K}}

\begin{document}

\title{Quasinormal modes and grey-body factors of Morris-Thorne wormholes}
\author{Zainab Malik}\email{zainabmalik8115@outlook.com}
\affiliation{Institute of Applied Sciences and Intelligent Systems, H-15, Pakistan}

\begin{abstract}
Using the fact that, for a broad class of Morris-Thorne wormholes, the maximum of the effective potential is located at the throat, we derive accurate analytic WKB expressions for the quasinormal modes and grey-body factors of various traversable wormholes. In the eikonal limit, these analytic expressions acquire a compact form and satisfy the correspondence between the quasinormal modes and the radii of the wormhole shadows.
\end{abstract}

\maketitle

\section{Introduction}

Traversable wormholes, originally introduced within the framework of general relativity \cite{Morris:1988cz}, represent hypothetical structures that connect two separate regions of spacetime. While their existence remains speculative, wormholes continue to capture the attention of researchers as solutions to Einstein's field equations that require exotic matter to sustain their geometry. Among various traversable wormhole models, those described by the Morris-Thorne ansatz have become particularly significant due to their simplicity and physical transparency.

The study of quasinormal modes (QNMs) and grey-body factors (GBFs) of wormholes has gained considerable interest in recent years \cite{DuttaRoy:2019hij,Azad:2022qqn,Blazquez-Salcedo:2018ipc,Bronnikov:2021liv,Churilova:2021tgn,Oliveira:2018oha,Batic:2024vsb,Ou:2021efv,Churilova:2019qph,Zhang:2023kzs,Azad:2023idq,Roy:2021jjg,Jusufi:2020mmy,Mitra:2023yjf,Gogoi:2022ove,Alfaro:2024tdr,Bronnikov:2019sbx}, as these quantities provide important insights into the spacetime's response to perturbations and its observational characteristics \cite{LIGOScientific:2016aoc,LIGOScientific:2017vwq,LIGOScientific:2020zkf,Oshita:2023cjz} . QNMs, representing damped oscillations of spacetime, are uniquely determined by the wormhole's geometry and boundary conditions \cite{Konoplya:2010kv}. Grey-body factors, on the other hand, quantify the transmission probability of waves propagating through the wormhole's effective potential barrier \cite{Konoplya:2010kv}. Together, these properties can serve as tools for distinguishing wormholes from black holes in gravitational wave astronomy and related observational studies. It is worth mentioning that the link between the two considered characteristics, quasinormal modes and grey-body factors, in the high frequency regime, have been recently disccussed in \cite{Oshita:2023cjz,Rosato:2024arw,Konoplya:2024lir,Oshita:2024fzf,Dubinsky:2024vbn,Skvortsova:2024msa} for black holes and extended in \cite{Bolokhov:2024otn} for wormholes.

In this paper, we study the QNMs and GBFs of several examples of traversable wormholes described by the Morris-Thorne metric. We focus on cases with spherically symmetric geometries, including both tideless and more general wormholes with non-zero tidal forces. By employing the WKB approximation  \cite{Schutz:1985km, Iyer:1986np, Konoplya:2003ii, Matyjasek:2017psv}, we compute the QNMs and GBFs.

The paper is organized as follows: In Sec.~II, we introduce the Morris-Thorne ansatz for wormhole spacetimes and describe the key properties of the metric and its components. Sec.~III focuses on quasinormal modes, detailing the use of the WKB method and its application to wormhole spacetimes. In Sec.~IV, we discuss grey-body factors and their computation using the WKB approach. Sec.~V provides analytic expressions for quasinormal modes and grey-body factors for several specific wormhole metrics, including examples with both tideless and non-tideless geometries. Finally, conclusions and potential extensions of this work are summarized in Sec.~VI.

\section{Morris-Thorne ansatz for wormholes' spacetimes}

Static, spherically symmetric, Lorentzian traversable wormholes, irrespective of their specific geometric configurations, can be elegantly described within the framework established by Morris and Thorne \cite{Morris:1988cz}. This class of spacetimes is characterized by the following metric:
\begin{equation}\label{MT}
ds^2 = - e^{2 \Phi(r)} dt^2 + \frac{dr^2}{1 - \frac{b(r)}{r}} + r^2 (d\theta^2 + \sin^2\theta \, d\phi^2),
\end{equation}
where the function $\Phi(r)$, often referred to as the lapse function, determines the redshift effects and tidal forces experienced within the wormhole spacetime. When $\Phi(r)$ remains constant throughout the spacetime, the wormhole is classified as "tideless," indicating that it does not exert tidal accelerations on point-like particles. However, as noted in \cite{Morris:1988cz}, while point particles may be unaffected, extended bodies traveling through the wormhole may still experience residual tidal forces depending on their structure and orientation. The overall geometry of the wormhole is governed by the function $b(r)$, commonly called the shape function, which determines the curvature of spacetime and ensures the wormhole's traversability.

The throat of the wormhole, which represents its narrowest cross-section, is located at the minimum value of the radial coordinate, denoted by $r_{\text{min}} = b_0$. The radial coordinate $r$ spans from this throat value to infinity on either side of the wormhole, with $r \to \infty$ corresponding to the asymptotically flat regions. To better analyze the geometric structure, it is convenient to introduce the proper radial distance $l$, defined through the relationship:
\begin{equation}
dl = \pm \left(1 - \frac{b(r)}{r}\right)^{-1/2} dr.
\end{equation}
In terms of this proper radial distance, the wormhole geometry exhibits symmetry, extending from $l = -\infty$ on one side of the throat to $l = +\infty$ on the other, with $r \to \infty$ corresponding to these asymptotic limits. This symmetry allows for a complete and continuous description of the wormhole's internal structure and external connections.

To ensure the spacetime is free of singularities, the lapse function $\Phi(r)$ must remain finite throughout the wormhole, particularly at the throat. Furthermore, the requirement of asymptotic flatness imposes the condition $\Phi(r) \to 0$ as $r \to \infty$, or equivalently, $l \to \pm \infty$. The shape function $b(r)$ must also satisfy specific criteria to maintain the physical traversability of the wormhole and ensure a non-pathological geometry. These include:
\begin{equation}
1 - \frac{b(r)}{r} > 0, 
\end{equation}
and, for asymptotically flat wormholes,
\begin{equation}
 \frac{b(r)}{r} \to 0 \quad \text{as} \quad r \to \infty.
\end{equation}
At the throat of the wormhole, where $r = b(r)$, the metric satisfies:
\begin{equation}
1 - \frac{b(r)}{r} \to 0.
\end{equation}
This condition ensures that the proper radial distance remains well-defined, preventing divergences in the metric components at the throat. The absence of singularities at the throat guarantees that traversable wormholes are, in principle, suitable for physical exploration, allowing matter or even hypothetical travelers to pass through the wormhole within a finite proper time.

The traversable wormhole metric described here is distinguished by its ability to connect distant regions of spacetime while maintaining a smooth and regular geometry at the throat. This makes such configurations invaluable for theoretical studies exploring the nature of spacetime topology and its potential applications, including interstellar travel and exotic communication channels. Furthermore, the conditions imposed on the lapse and shape functions serve as essential constraints for the construction of physically viable wormhole models that comply with general relativity and its extensions.

\section{Quasinormal modes}

After separating variables, the perturbation equations for test scalar and electromagnetic fields can be rewritten in the Schrödinger wave-like form:
\begin{equation}\label{sp}
\frac{d^2\Psi}{dr_*^2} + \omega^2 \Psi - V(r) \Psi = 0,
\end{equation}
in terms of the tortoise coordinate, $r_*$, is defined by the following equation,
\begin{equation}
dr_* = \pm \frac{dr}{f(r)},
\end{equation}
where
\begin{equation}
f(r) =\sqrt{1 - \frac{b(r)}{r}}.
\end{equation}
The effective potential $V(r)$ is given by:
\begin{equation}
V(r) = \frac{e^{2\Phi}}{r^2} \ell (\ell+1) +  \frac{1}{2r} \frac{d}{dr} \left( f(r)^2 \right),
\end{equation}
for scalar perturbations, and, 
\begin{equation}
V(r) = \frac{e^{2\Phi}}{r^2} \ell (\ell+1),
\end{equation}
for the electromagnetic ones. 
The effective potentials have their maximum value located at the throat of the wormhole $r=b_0$ and monotonically decay at both infinities, representing either two different universes or two distant regions within the same universe. 

Therefore, it is natural that the quasinormal modes of wormholes satisfy the following boundary conditions as those for black holes \cite{Konoplya:2005et}:
\begin{equation}
\Psi \sim e^{\pm i \omega r_*}, \quad r_* \rightarrow \pm \infty.
\end{equation}
These conditions imply purely outgoing waves at both infitinies.

The higher order WKB method we use \cite{Schutz:1985km, Iyer:1986np, Konoplya:2003ii, Matyjasek:2017psv} implies that there are two turning points and a monotonically decreasing effective potential towards both infinities in terms of the tortoise coordinate. Then, 
the relation allowing us to find the quasinormal modes is
\begin{equation}\label{WKB}
\frac{i Q_{0}}{\sqrt{2 Q_{0}''}} - \sum_{i=2}^{p} \Lambda_{i} - n - \frac{1}{2}=0, 
\end{equation}
where $\Lambda_i$ are correction terms, derived at various orders in \cite{Schutz:1985km, Iyer:1986np, Konoplya:2003ii, Matyjasek:2017psv} which depend on the derivatives of the effective potential at the maximum up to the order $2i$. Here, $Q = \omega^2 - V$ and $Q_0^{i}$ is the $i$-th derivative of it at the potential's peak, 
$n$ is the overtone number.

The WKB approach has been used for finding quasinormal modes of black holes and wormholes in a great number of publications \cite{Zinhailo:2019rwd, Dubinsky:2024fvi, Konoplya:2005sy, Skvortsova:2024wly, Guo:2023ivz, Bolokhov:2023dxq, Paul:2023eep, Cuyubamba:2016cug, Kokkotas:2010zd, Gong:2023ghh, Zhidenko:2008fp, Zinhailo:2018ska, Skvortsova:2024atk, Malik:2024tuf, Malik:2024bmp, Bolokhov:2023bwm, Dubinsky:2024hmn,  Konoplya:2001ji}, showing good concordance with other other method for the low-lying frequencies. 

\section{Grey-body factors}

The grey-body factors are eseentially related to the transmission coefficients for a portion of radiation which penetrates the potential barrier and reach a distant observer. Owing to the symmetry of the problem, the following boundary conditions are imposed for this scattering process:
\begin{equation}
\begin{array}{rclcl}
\Psi &=& e^{-i\Omega r_*} + R e^{i\Omega r_*}, &\quad& r_* \to +\infty, \\
\Psi &=& T e^{-i\Omega r_*}, &\quad& r_* \to -\infty,
\end{array}
\end{equation}
where $R$ is the reflection coefficient, and $T$ is the transmission coefficient. Following recent publications \cite{Konoplya:2024lir} we distinguish the continuous and purely real frequency $\Omega$ for the scattering problem from the complex and quasinormal mode $\omega_{n}$ running a set of discrete values.  
The grey-body factor is then defined as follows:
\begin{equation}
\Gamma_{\ell}(\Omega) = |T|^2 = 1 - |R|^2.
\end{equation}
The WKB formula uses the following ansatz for the grey-body factors
\begin{equation}\label{eq:gbfactor}
\Gamma_{\ell}(\Omega) = \frac{1}{1 + e^{2\pi \imo \K}}.
\end{equation}
The latter relation was used for finding grey-body factors in a number of recent works  \cite{Konoplya:2023ahd, Dubinsky:2024nzo, Toshmatov:2015wga, Bolokhov:2024voa, Skvortsova:2024msa, Konoplya:2020jgt}.

Notice that, when deriving analytic expressions for quasinormal modes and grey-body factors using the WKB approach, we employ two slightly different methods to address the problems. To calculate the quasinormal modes, we expand the WKB formula into a series in powers of $\kappa = (\ell + 1/2)^{-1}$, in similar way with \cite{Konoplya:2023moy,Malik:2024tuf,Malik:2024bmp,Malik:2024voy,Malik:2024sxv}. However, when determining the quantity $\mathcal{K}$, which governs the grey-body factors, we avoid further expansion in terms of the inverse of the multipole number. This way we find a balance between the compactness and accuracy of the resultant expressions. For both characteristics we are able to obtain more accurate expressions than in the aforementioned case of black holes, becuase the maximum of the peak of the effective potential is known exactly and does require further expansion in terms of the inverse multipole number.

It should be noted that the WKB method may fail to describe certain configurations, even in the regime of high multipole numbers where it is expected to be accurate. This can occur when the effective potential exhibits an unusual centrifugal barrier that deviates from the standard form $f(r) \ell (\ell + 1)/r^2$. For instance, in theories with higher curvature corrections, the perturbation may even become unstable (see, for example, \cite{Takahashi:2011du,Takahashi:2010gz,Gleiser:2005ra,Dotti:2005sq,Konoplya:2017ymp,Takahashi:2011qda,Konoplya:2017lhs,Konoplya:2020bxa}). Cases where the WKB method breaks down in the eikonal limit or is incomplete are summarized in \cite{Khanna:2016yow,Konoplya:2019hml,Bolokhov:2023dxq}.

\section{Applications to various wormhole metrics}

In this section, we derive analytic expressions for the quasinormal modes and grey-body factors of various wormhole spacetimes using the 6th-order WKB formula. For certain types of metrics that allow for a Taylor series expansion near the throat, analytic expressions for quasinormal modes were obtained in the dominant orders in \cite{Konoplya:2018ala}. However, the asymptotic behavior of such wormholes is generally not asymptotically flat, and these solutions are typically matched with Schwarzschild-like geometries at a finite distance from the throat.

\vspace{3mm}
\subsection{Morris-Thorne Wormhole type 1: Ellis-Bronnikov solution}

One of the very well-known examples is tideless Ellis-Bronnikov wormhole \cite{Bronnikov:1973fh,Ellis:1973yv} for which:
\begin{equation}
\Phi(r) = 0, 
\end{equation}
\begin{equation}
b(r) = \frac{b_{0}^2}{r}.
\end{equation}
Thus, the maximum effective potential is located at $r =b_{0}$. Here we measure all quantities in units $b_{0}=1$.
Using the WKB formula at the 6th order we obtain the expression for quasinormal modes for the \textit{electromagnetic perturbations}:
\begin{widetext}
\begin{align}
\omega = & \, \kappa - i \mathcal{K} 
+ \frac{-4 \mathcal{K}^2 - 5}{16 \kappa} 
- \frac{i \mathcal{K} \left(4 \mathcal{K}^2 - 21\right)}{64 \kappa^2} 
+ \frac{-16 \mathcal{K}^4 + 184 \mathcal{K}^2 + 179}{2048 \kappa^3} \notag \\
& - \frac{3 i \mathcal{K} \left(80 \mathcal{K}^4 - 120 \mathcal{K}^2 + 1961\right)}{16384 \kappa^4} 
+ \frac{288 \mathcal{K}^6 + 288 \mathcal{K}^4 - 36654 \mathcal{K}^2 - 16147}{65536 \kappa^5} 
+ O\left(\frac{1}{\kappa}\right)^6.
\end{align}
Here we used $\K = n+\frac{1}{2}$ and $\kappa = \ell +\frac{1}{2}$.
The greybody factors can be found from eq. \ref{eq:gbfactor}, where $\K$ is no longer $n+(1/2)$, but can be found from the following relation:
\begin{small}
\begin{align}
-\frac{1048576 \kappa^{13}}{i} \mathcal{K} = & \, 
    1111680 \kappa^{14} 
    - 9 \kappa^{12} \left(299260 \Omega^2 + 22881\right)  + 3 \kappa^{10} \Big(1240600 \Omega^4 - 59101 \Omega^2 + 28906 \Big) \notag \\
& - \kappa^8 \Big(4084740 \Omega^6 - 91958 \Omega^4 + 438606 \Omega^2 + 626201 \Big) + \kappa^6 \Omega^2 \Big(3049200 \Omega^6 - 51710 \Omega^4 + 514946 \Omega^2 + 475981 \Big) \notag \\
& + \kappa^4 \Omega^6 \Big(-1460004 \Omega^4 + 18115 \Omega^2 - 153330 \Big) + \kappa^2 \Omega^{10} \Big(404936 \Omega^2 - 2811 \Big) 
    - 49532 \Omega^{14}.
\end{align}
\end{small}
For \textit{scalar perturbations} we find for quasinormal modes,
\begin{align}
\omega = & \, \kappa - i \mathcal{K} 
+ \frac{3 - 4 \mathcal{K}^2}{16 \kappa} 
- \frac{i \mathcal{K} \left(4 \mathcal{K}^2 + 11\right)}{64 \kappa^2} 
+ \frac{-16 \mathcal{K}^4 - 72 \mathcal{K}^2 - 141}{2048 \kappa^3} \notag \\
& - \frac{i \mathcal{K} \left(240 \mathcal{K}^4 + 152 \mathcal{K}^2 - 4997\right)}{16384 \kappa^4} 
+ \frac{288 \mathcal{K}^6 + 32 \mathcal{K}^4 + 31826 \mathcal{K}^2 + 13341}{65536 \kappa^5} 
+ O\left(\frac{1}{\kappa}\right)^6,
\end{align}
and the quantity $\K$, can be found from the following relation
\begin{align}
-\frac{1048576 \kappa^{13}}{i} \mathcal{K} = & \,
1111680 \kappa^{14} 
- 45 \kappa^{12} \left(59852 \Omega^2 - 3179\right) + \kappa^{10} \Big(3721800 \Omega^4 + 75121 \Omega^2 - 115282\Big) \notag \\
& + \kappa^8 \Big(-4084740 \Omega^6 - 35098 \Omega^4 + 313378 \Omega^2 + 298975\Big) \notag \\
& + \kappa^6 \Omega^2 \Big(3049200 \Omega^6 + 19090 \Omega^4 - 421934 \Omega^2 - 16267\Big) \notag \\
& + \kappa^4 \Omega^6 \Big(-1460004 \Omega^4 - 6565 \Omega^2 + 127070\Big)  + \kappa^2 \Omega^{10} \Big(404936 \Omega^2 + 1005\Big) 
- 49532 \Omega^{14}.
\end{align}

\subsection{Morris-Thorne Wormhole type 2}

One more example is the tideless Morris-Thorne wormhole for which the shape function is given by \cite{Morris:1988cz}.:
\begin{equation}\label{MT2}
b(r) = \sqrt{b_{0} r}.
\end{equation}

Quasinormal modes for \textit{electromagnetic perturbations} are 
\begin{align}
\omega = & \, \kappa 
- \frac{i \mathcal{K}}{2} 
- \frac{5 \left(4 \mathcal{K}^2 + 17\right)}{512 \kappa} 
- \frac{5 i \mathcal{K} \left(52 \mathcal{K}^2 - 137\right)}{32768 \kappa^2} \notag \\
& + \frac{5 \left(3792 \mathcal{K}^4 - 23736 \mathcal{K}^2 - 35791\right)}{16777216 \kappa^3} 
+ \frac{i \mathcal{K} \left(16912 \mathcal{K}^4 - 3757720 \mathcal{K}^2 + 9918933\right)}{2147483648 \kappa^4} \notag \\
& + \frac{8326304 \mathcal{K}^6 + 24282400 \mathcal{K}^4 - 184256694 \mathcal{K}^2 - 145331135}{137438953472 \kappa^5} 
+ O\left(\frac{1}{\kappa}\right)^6.
\end{align}
While the grey-body factors can be found from $\K$ of the following form
\begin{align}
-\frac{8796093022208 \kappa^{13}}{i} \mathcal{K} = & \,
16797975744256 \kappa^{14} 
- 16 \kappa^{12} \Big(2345776771856 \Omega^2 + 197510710155\Big) \notag \\
& + 24 \kappa^{10} \Big(1993038958368 \Omega^4 + 23380579770 \Omega^2 - 9104769121\Big) \notag \\
& - \kappa^8 \Big(51380539540736 \Omega^6 + 592577752800 \Omega^4 + 28218625080 \Omega^2 + 39702583535\Big) \notag \\
& + \kappa^6 \Omega^2 \Big(37974779178240 \Omega^6 + 395027073760 \Omega^4 + 19108746168 \Omega^2 + 5087105295\Big) \notag \\
& - 24 \kappa^4 \Omega^6 \Big(753280813600 \Omega^4 + 6144676730 \Omega^2 + 265443311\Big) \notag \\
& + 48 \kappa^2 \Omega^{10} \Big(104072033072 \Omega^2 + 489220985\Big) 
- 609440093952 \Omega^{14}.
\end{align}
The quasinormal modes for \textit{scalar perturbations} are
\begin{align}
\omega = & \, \kappa 
- \frac{i \mathcal{K}}{2} 
+ \frac{-20 \mathcal{K}^2 - 21}{512 \kappa} 
- \frac{i \mathcal{K} \left(260 \mathcal{K}^2 - 173\right)}{32768 \kappa^2} 
+ \frac{18960 \mathcal{K}^4 - 28568 \mathcal{K}^2 + 7413}{16777216 \kappa^3} \notag \\
& + \frac{i \mathcal{K} \left(16912 \mathcal{K}^4 - 898712 \mathcal{K}^2 - 1535531\right)}{2147483648 \kappa^4} \notag \\
& + \frac{8326304 \mathcal{K}^6 + 5948704 \mathcal{K}^4 - 22931638 \mathcal{K}^2 - 11190207}{137438953472 \kappa^5} 
+ O\left(\frac{1}{\kappa}\right)^6.
\end{align}
and the grey-body factors are found from $\K$ which have the same form as for electromagnetic perturbations.

\subsection{Morris-Thorne Wormhole type 3}

Now, we consider a more general case of a Morris-Thorne wormhole where the tidal forces at the throat are non-zero. This occurs when the redshift function is no longer constant. The shape and redshift functions for this wormhole are given by:
\begin{equation}\label{MT2}
b(r) = b_0 \left(\frac{b_0}{r}\right)^q, \; q < 1, \quad \Phi(r) = \frac{1}{r^p}, \; p \geq 0.
\end{equation}
The scattering properties and quasinormal ringing of this wormhole have been extensively studied in \cite{Konoplya:2010kv}.
We will concentrate here in the tideless case $ \Phi(r) =0$. For \textit{electromagnetic perturbations} we have 
\begin{align}
\omega = & \, \kappa 
- \frac{i}{2} \sqrt{2 - 2q} \, \mathcal{K} 
+ \frac{q^2 \Big(-\big(4 \mathcal{K}^2 + 1\big)\Big) + 4q \big(4 \mathcal{K}^2 + 3\big) - 3\big(4 \mathcal{K}^2 + 9\big)}{128 \kappa} \notag \\
& + \frac{i (q-3)(q-1)^2 \mathcal{K} \Big(q \big(20 \mathcal{K}^2 + 31\big) + 68 \mathcal{K}^2 - 221\Big)}{6144 \sqrt{2 - 2q} \kappa^2} \notag \\
& - \frac{(q-3)}{1572864 \kappa^3} 
\Big(15 q^4 \big(16 \mathcal{K}^4 + 104 \mathcal{K}^2 + 21\big) 
+ 32 q^3 \big(64 \mathcal{K}^4 - 436 \mathcal{K}^2 - 129\big) \notag \\
& \quad + q^2 \big(-3296 \mathcal{K}^4 + 22736 \mathcal{K}^2 + 15930\big) 
- 128q \big(4 \mathcal{K}^4 + 29 \mathcal{K}^2 + 126\big) 
+ 1520 \mathcal{K}^4 - 6632 \mathcal{K}^2 - 2133\Big) \notag \\
& + O\left(\frac{1}{\kappa}\right)^4.
\end{align}
and for the grey-body factors 
\begin{small}
\begin{align}
-\frac{37748736 \sqrt{2 - 2q} \kappa^9}{i} \mathcal{K} = & \,
12 \Big(7 q^4 + 268 q^3 + 2522 q^2 - 326740 q + 5692887\Big) \kappa^{10} 
- 12 \Big(7 q^4 + 188 q^3 + 298 q^2 - 64452 q + 357063\Big) \Omega^{10} \notag \\
& + 4 \kappa^8 \Big(
    1000 q^5 
    - 5 q^4 \big(21 \Omega^2 + 4936\big) 
    - 140 q^3 \big(27 \Omega^2 - 808\big) 
    - 62 q^2 \big(453 \Omega^2 - 1360\big) \notag \\
& \quad + 12 q \big(271685 \Omega^2 + 42218\big) 
    - 423 \big(77167 \Omega^2 + 8456\big)
\Big) \notag \\
& + 4 \kappa^2 \Omega^6 \Big(
    -1000 q^5 
    + 5 q^4 \big(21 \Omega^2 + 3928\big) 
    + 20 q^3 \big(153 \Omega^2 - 5728\big) 
    + 10 q^2 \big(807 \Omega^2 + 17776\big) \notag \\
& \quad - 12 q \big(99545 \Omega^2 - 5902\big) 
    + 9 \big(748625 \Omega^2 - 7448\big)
\Big) \notag \\
& + \kappa^6 \Big(
    2527 q^6 
    - 6 q^5 \big(2000 \Omega^2 + 12291\big) 
    + q^4 \big(840 \Omega^4 + 276000 \Omega^2 + 734737\big) \notag \\
& \quad + 12 q^3 \big(2360 \Omega^4 - 126400 \Omega^2 - 264013\big) 
    + q^2 \big(159024 \Omega^4 + 423552 \Omega^2 + 5836417\big) \notag \\
& \quad - 6 q \big(2925072 \Omega^4 - 1117648 \Omega^2 + 524131\big) 
    + 9 \big(12479880 \Omega^4 - 268448 \Omega^2 - 109865\big)
\Big) \notag \\
& - \kappa^4 \Omega^2 \Big(
    2527 q^6 
    - 6 q^5 \big(2000 \Omega^2 + 10219\big) 
    + q^4 \big(840 \Omega^4 + 255840 \Omega^2 + 552289\big) \notag \\
& \quad + 12 q^3 \big(2200 \Omega^4 - 126880 \Omega^2 - 191909\big) 
    + q^2 \big(105648 \Omega^4 + 2061696 \Omega^2 + 4494049\big) \notag \\
& \quad - 6 q \big(2072528 \Omega^4 - 322960 \Omega^2 + 575179\big) 
    + 9 \big(8081416 \Omega^4 - 118496 \Omega^2 + 45223\big)
\Big).
\end{align}
\end{small}
When the tidal force is non-zero at the throat and $\Phi(r) = \frac{1}{r^p}$, we have the following expression for the quasinormal modes 
\begin{align}
\omega = & \, e \kappa 
- \frac{i e \mathcal{K} \sqrt{(p+1)(1-q)}}{\sqrt{2}} \notag \\
& + \frac{e}{128 \kappa} \Bigg(
    -\frac{(q-1) \left(4 \mathcal{K}^2 + 1\right) \big(-13 p^2 + p (q - 21) + q - 11\big)}{p+1} 
    - 32 (p+1)(q-1) \mathcal{K}^2 
    - 16
\Bigg) \notag \\
& + \frac{i e (1-q)^{3/2} \mathcal{K}}{6144 \sqrt{2} (p+1)^{5/2} \kappa^2} 
\Bigg(
    p^4 \big(1135 - 364 \mathcal{K}^2\big) 
    + p^3 \Big(6 q \big(4 \mathcal{K}^2 - 53\big) - 856 \mathcal{K}^2 + 3550\Big) \notag \\
& \quad + p^2 \Big(q^2 \big(20 \mathcal{K}^2 + 31\big) 
        + q \big(80 \mathcal{K}^2 - 884\big) 
        - 1300 \mathcal{K}^2 + 3889\Big) \notag \\
& \quad + p \Big(q^2 \big(40 \mathcal{K}^2 + 62\big) 
        + 16 q \big(4 \mathcal{K}^2 - 55\big) 
        - 808 \mathcal{K}^2 + 2338\Big) \notag \\
& \quad + (q-3) \Big(q \big(20 \mathcal{K}^2 + 31\big) 
        + 68 \mathcal{K}^2 - 221\Big)
\Bigg) 
+ O\left(\frac{1}{\kappa}\right)^3.
\end{align}
\end{widetext}

The expressions for gray-body factors are much more lengthy and, therefore, are not written here. In principle, the WKB approach could be effectively used to find analytic expressions of quasinormal modes and graybody factors of various traversable wormholes, once the peak of the effective potential coincides with the radius of the throat. 

\subsection{Analytic expressions in the eikonal limit}

In the eikonal limit $\ell \rightarrow \infty$ the expressions for quasinormal modes and $\K$ become especially simple and usually do not depend on the spin of the perturbed field.
Here we will keep the explicit dependence of the quasinormal modes on the overtone number $n$ and of all quantities on the throat radius $b_0$. 

For the \textit{first Morris-Thorne} model we have,
\begin{equation}\nonumber
\omega_{n} b_0 =  \kappa -i (n+(1/2) +O\left(\frac{1}{\kappa }\right),
\end{equation}
\begin{equation}\nonumber
\K = \frac{i \left(b_{0}^{2} \Omega ^2-\kappa ^2\right)}{2 \kappa}.
\end{equation}

For the \textit{second Morris-Thorne model} we find
\begin{equation}\nonumber
\omega_{n} b_0 =  \kappa -\frac{i (n+(1/2))}{2}+O\left(\frac{1}{\kappa
   }\right),
\end{equation}
\begin{equation}\nonumber
\K =  \frac{i \left(b_{0}^{2}  \Omega ^2-\kappa ^2\right)}{\kappa}.
\end{equation}

For the \textit{third Morris-Thorne model} we obtain 
\begin{equation}\nonumber
\omega_{n} b_0 = e \kappa -\frac{i e (n+(1/2)) \sqrt{(p+1)
   (q+1)}}{\sqrt{2}}+O\left(\frac{1}{\kappa
   }\right)
\end{equation}

\begin{equation}\nonumber
\K = \frac{i \left(b_{0}^{2} \Omega ^2-e^2 \kappa ^2\right)}{\sqrt{2} e^2
   \kappa  \sqrt{(p+1) (q+1)}}, \quad \kappa =\ell +\frac{1}{2}
\end{equation}

Using the above relations one can immediately show that the correspondence between the eikonal quasinormal modes and the radii of shadows cast  by the wormholes \cite{Jusufi:2020mmy} is satisfied for the considered cases. Indeed
the correspondence says that
\begin{equation}
\omega_R = \frac{1}{R_s} \left(\ell + \frac{1}{2} \right). \tag{29}
\end{equation}

Using the equation for the radius of the shadow:
\begin{equation}
R_s = \sqrt{\alpha^2 + \beta^2} = \frac{b_0}{\Phi(b_0)}, \tag{22}
\end{equation}
where we have 
\begin{equation}
\alpha = \lim_{r \to \infty} \left( -r^2 \sin^2 \theta_0 \frac{d\phi}{dr} \right), \tag{18}
\end{equation}
and
\begin{equation}
\beta = \lim_{r \to \infty} \left( r^2 \frac{d\theta}{dr} \right), \tag{19}
\end{equation}
one can easily check that the obtained eikonal expressions for the quasinormal modes satisfy the above correspondence with the radius of the shadow.

\vspace{3mm}
\section{Conclusions}

Here we have found analytic expressions for quasinormal modes and gray-body factors of various types of traversable wormhole within the Morris-Thorne ansatz. Relatively simple analytic expressions are obtained because the peak of the effective potential for the considered class of wormholes and fields is located exactly on the throat of the wormhole. 
We have shown that in the eikonal limit the obtained analytic expressions for quasinormal frequencies satisfy the correspondence with the radii of shadows cast by wormholes.  

This approach could be extended to rotating wormholes in a similar manner. 
In this case, in some range of parameters rotating wormholes may have a phenomenon of superradiance, which is well known for axially symmetric black holes \cite{1971JETPL..14..180Z,Starobinskil:1974nkd,Starobinsky:1973aij,Bekenstein:1998nt,East:2017ovw,Hod:2012zza,Konoplya:2008hj} or charged fields in the background of charged black holes \cite{Bekenstein:1998nt,Konoplya:2014lha,Richarte:2021fbi,Zhu:2014sya}. The effect of superradiance provides that the reflection coefficient is greater than unity, which is obviously outside the scope of applicability of the WKB method.

\acknowledgments{
The author acknowledges Roman Konoplya for useful and encouraging discussions.

\bibliography{bibliography}

\end{document}